# RAVEN-X: A High Performance Data Mining Toolbox for Bioacoustic Data Analysis

Peter J. Dugan and Holger Klinck
Bioacoustics Research Program, Cornell Lab of Ornithology, Cornell University
159 Sapsucker Woods Road, Ithaca, NY 14850
phone: (607) 254-1149    fax: (607) 254-2460    email: pjd78@cornell.edu

Marie A. Roch
Department of Computer Science, San Diego State University
5500 Campanile Drive, San Diego, CA 92182

Tyler A. Helble
US Navy, SPAWAR Systems Center Pacific
53560 Hull Street, San Diego, CA 92152

## LONG TERM GOALS

The goals of this project are [a] to develop a MATLAB toolbox, called Raven-X, intended for high performance data mining (detection and classification of target signals) of BIG acoustic datasets and [b] to make the toolbox freely available to the bioacoustic community.

## OBJECTIVES

Our objective is to integrate high performance computing (HPC) technologies and bioacoustics data-mining capabilities by offering a MATLAB-based toolbox called Raven-X. Raven-X will provide a hardware-independent solution, for processing large acoustic datasets - the toolkit will be available to the community at no cost. This goal will be achieved by leveraging prior work done by Dugan, et al., [1-3] which successfully deployed MATLAB based HPC tools within Cornell University's Bioacoustics Research Program (BRP). These tools enabled commonly available multi-core computers to process data at accelerated rates to detect and classify whale sounds in large multi-channel sound archives [4, 5]. Through this collaboration, we will expand on this effort which was featured through Mathworks research and industry forums [6, 7], incorporate new cutting-edge detectors and classifiers, and disseminate Raven-X to the broader bioacoustics community.

## APPROACH

Raven-X will provide capabilities to accelerate processing of large sound archives. The toolkit is designed to run on commercially available off the shelf (COTS) hardware (e.g. standard computers) as well as HPC technologies (e.g. cloud computers with multiple nodes). Raven-X will be designed to handle various sound formats; (e.g. flac, wav, aif, and M3R dat) and recording scenarios; such as multi-channel data and intermittent recordings. The goal is to make this tool accessible and useful for running a host of algorithms.

The project is broken down into two phases of equal duration. Phase-I we aim to integrate existing HPC models [1, 2, 8] and provide a stable software tool through an online repository. The toolkit will be deployed to run on standard computers as well as HPC technologies (e.g. cloud computers). An HPC model [1] will be added to provide accelerated processing. Popular algorithms will be integrated to the toolkit, these include GPL [9], silbido [10] and ERMA [11]. The team will work through various architectural issues and changes will be incorporated into a fault tolerant application. Each algorithm will be tested and benchmarked under various conditions, such as sample rate, archive size, channel configuration and sound file type.

Phase-II will focus on creating a deployable package to the broader community. We aim to create and modify new user interfaces, documentation will be added where necessary. Interfaces will be open, allowing other software experts, from the community, to explore adding custom algorithms. For deployed algorithms, Raven-X will enable users to export detector/classifier results in a format which can be used by existing software tools (e.g. Raven Pro and Tethys). The objective is to build a stable package, robust and capable of running a series of new algorithms for advanced marine mammal detection and classification. At the end of the project, Raven-X will be made available to the bioacoustics community at no-cost through an online repository.

## WORK COMPLETED

### Computer Systems Research
*Investigating Hardware for PMRF*

Tyler Helble investigated several hardware configurations for improving computing performance at SSC-PAC processing for PMRF and SCORE data. Tyler consulted with the Cornell team, SSC-PAC's High Performance Computing team, and attended the 2016 passive acoustic data archiving meeting at NOAA in Boulder, CO. It was determined that a lack of high speed, online storage is a main bottleneck for processing range data. A 270TB network attached storage (NAS) device was purchased with end-of-year PACFLEET funding. The unit is manufactured by NEXSAN BEAST storage; installation for SSC-PAC lab system is scheduled for the end of October. The storage device is expandable and will allow the lab to host nearly all of the unclassified acoustic data on a single accessible device, rather than stored on hundreds of unconnected SATA drives. This will eliminate the need for an operator to physical swap drives, and will allow for a more automated processing of PMRF data; minimizing human intervention. The acquisition of the NAS, combined with improvements in algorithm speeds from this project, should allow SSC-PAC to process each year's worth of data significantly faster than traditional methods[1]. The NAS will be hosted in the SSC-PAC lab and connected to an existing 16 core machine. The NAS device is also compatible with the SSC-PAC's, 144 core, 40TB RAM system located in the HPC center, and may eventually be migrated to that system. The connectivity to the HPC is relatively slow (1 gig Ethernet), so the lab may need to be physically relocated to an area with faster network speeds for the data to be hosted at the HPC center. Additionally, any computer connected to the HPC must be on the SSC-PAC network, which means that several of our custom software packages for processing data would need to be added to DATUMS before they are cleared for use on the SSC-PAC network. Therefore, we foresee the migration to the HPC to be at least several years down the road.

---

[1] Using the new system we estimate processing time to be on the order of a week - versus the current configuration which takes months.

*Raven-X Software*

HPC runtime models, for the parallel-distributed processing, were updated and integrated into the Raven-X toolkit. Older user interfaces have been redesigned and implemented using the new object oriented (OO) structure, released in MATLAB 2016a. The goal is to create a complete OO toolkit, maximizing the ability to use components in Raven-X as building blocks for other HPC applications. The new Raven-X APP is shown in figure 1.

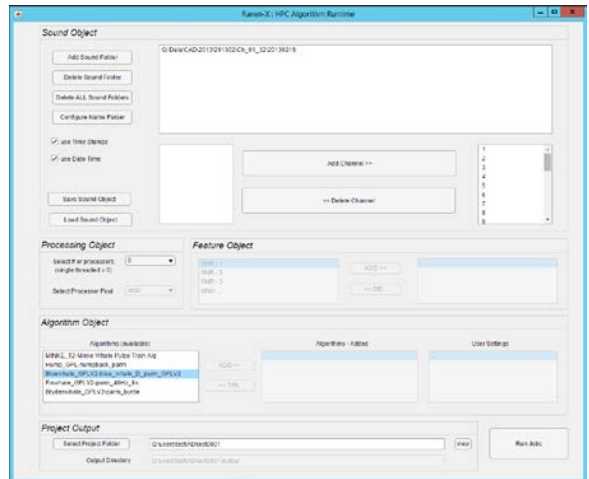

*Figure 1. Raven-X APP for running algorithms on large datasets. APP, OO-based UI tool, is built using the Mathworks APP Designer (new in 2016a).*

**Algorithm Interoperability**

*Algorithm Integration*

Four main algorithms, internal to Cornell BRP were integrated and tested in the new OO framework. These include three right whale algorithms and one algorithm designed for minke-whale detection. Preliminary integration of the GPL algorithm was accomplished using the OO framework. GPL comes equipped with various parameter files (called *parm-files*), each *parm-file* contains a series of settings that are used inside the GPL code. The OO structure provided a wrapper which allowed the original GPL package to remain untouched. This is best illustrated in figure 2. Integration with silbido and ERMA will continue after GPL integration is complete.

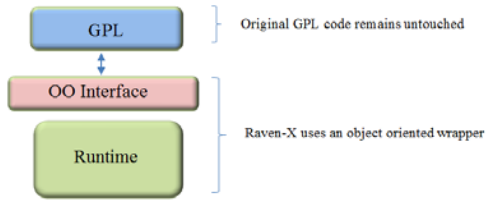

*Figure 2. By providing an object oriented interface, Raven-X incorporates advanced algorithms without changing any of the native code inside GPL. This makes both GPL and Raven-X independent from each other thereby incorporating future changes is more seamless and easier.*

*Algorithm Runtime Performance*
The GPL detector was successfully integrated into the Raven-X framework. Runtime performance was measured by processing GPL blue whale and GPL fin whale detector across 24 hours of single channel, 2 kHz data. The COTS computer selected for the test was equipped with 12 CPU's. The test consisted of two methods. *Method one* used the tradition approach to process the data for fin and blue whale detectors. This approach, referred to as the *Karoline Script*, was only capable of using 1 of the CPU's, leaving the other 11 CPU's idle. Results for *method one* shown in Table I below. The second method, *method two*, used the Raven-X software to run the exact same blue whale and fin whale algorithms. Raven-X was able to take advantage of all 12 CPU's; results for *method two* summarized in Table II.

| *Table I. Method One: Running GPL Algorithm using Karoline Script* | | | | |
|---|---|---|---|---|
| Number of Cores | Number of Cores | Execution Run Time (ERT) (seconds) | Number of Events | Rate (x Real Time = ERT/3600) |
| Blue Whale | 1 | 450.2 | 1194 | 8.0 |
| Fin Whale | 1 | 314.7 | 617 | 11.5 |

| *Table II. Method Two: Running GPL Algorithm using Raven-X HPC software* | | | | |
|---|---|---|---|---|
| Number of Cores | Number of Cores | Execution Run Time (ERT) (seconds) | Number of Events | Rate (x Real Time = ERT/3600) |
| Blue Whale | 12 | 76.3 | 1194 | 47.1 |
| Fin Whale | 12 | 56.4 | 617 | 63.8 |

**Data Formats and Fault Tolerance**
*Homogenous and Heterogeneous Data*
Homogenous detector outputs were successfully provided and tested for the Raven Pro format. This was tested for single and multi-channel cases. Future work will consider additional output formats (e.g. Tethys); these formats may require heterogeneous data. Initial work with heterogeneous data was done using a relatively sparse database containing programmatic and deployment metadata. This work is ongoing.

*Fault Tolerance, Algorithm Pre-Configuration*
Two sources of error were discovered during the initial phase of integrating GPL. The first deals with data integrity. The Raven-X runtime engine provides a mapping of sound data, adjusting for errors and drop out conditions; in some cases this is handled using a zero-padding technique. The GPL code requires only valid data (e.g. non-zero data), otherwise fatal errors inside GPL may occur. The second source of error was caused by incorrect configuration data. Since most of the algorithms rely on spectrogram based signal detection, each algorithm may require a specific set of parameters for proper execution. Solutions for errors are currently being addressed. First, the team will consider various methods to check data integrity prior to invoking the detection-classification algorithm(s); more robust error handling in the algorithm is also desirable. Avoiding incorrect configuration data will require a pre-setup routine that will be responsible for establishing proper combinations of parametric data. These routines will be specific to the algorithm; Raven-X will add necessary methods for calling these from the wrapper layer.

**RESULTS**

Work during the initial phase achieved several results. Study was done to investigate the upgrading the SSC-PAC computer system. A 270 TB NAS unit was purchased, allowing for PMRF and SCORE data to be processed significantly faster. Raven-X software was integrated from earlier working HPC models and software runtime was improved and fitted with a new OO based user interface. GPL algorithm was successfully integrated using a mid-level wrapper; the technique provided an integration point between Raven-X and GPL packages. Similar strategies will be used with ERMA and silbido. Fin whale and Blue whale GPL detectors were tested using a 24 hour, 2 kHz single channel data. Two different methods were tested; *method one* showing runtime using the traditional code and *method two* using Raven-X. *Method one* execution time for blue whale detector was 450.2 seconds versus 76.3 seconds using Raven-X (*method two*). Fin whale results were similar in comparison with an execution time of 314.7 seconds (*method one*) and 56.4 seconds using Raven-X (*method two*). Homogenous datatypes were successful at producing output formats for Raven Pro. Preliminary work with heterogeneous data was accomplished using a test database, significant progress will be addressed in future work along with integration into the Tethys database.

**IMPACT/APPLICATIONS**

U.S. Navy funded data collection and research efforts often result in large sound archives with mixtures of file formats and metadata. The project team is at the forefront of this issue. SPAWAR analyses multi-channel, high-frequency M3R data sets collected at various Navy training ranges. SDSU works closely with the Whale Acoustics Lab at SIO, which utilizes HARPs to monitor beaked whales and other cetaceans at numerous locations in U.S. waters. Analyzing these data sets is typically achieved by using a post-processing, ad hoc strategy, whereby various algorithms are tuned to extract cetacean calls of interest. Post-processing ideally runs as fast as possible, and it is not uncommon for processing rates to far exceed real-time rates by orders of magnitude. High rates for archival systems are achieved through processing many files in parallel (distributed computing). Higher rates are further achieved by scaling to larger computer systems for increased throughput. Raven-X will provide the bioacoustics community with a new and freely available tool to analyze large existing sound archives in a parallel mode. The toolbox will maximize algorithm speed and minimize analysis time and thus benefit future analysis efforts funded by the Navy.

**RELATED PROJECTS**

None at this time.